\RequirePackage{lineno}
\documentclass[aps,twocolumn,superscriptaddress,showpacs,amsmath,amssymb,nofootinbib]{revtex4}
 \pdfoutput=1
 \usepackage{graphicx,epsf,amsmath}
 \usepackage[]{latexsym}
 \usepackage{amsmath}
 \usepackage{subcaption}
 \usepackage{float}
 \usepackage{color}
 
 \usepackage{dcolumn}   
 \usepackage{bm}        
 \usepackage{amssymb}   

 \newcommand{\be}{\begin{eqnarray}}
 \newcommand{\ee}{\end{eqnarray}}

\begin{document}
 
 
\title{Simulation study of the correlation ($X_{max}^{\mu}$, $N^{\mu}$) in view of obtaining information on primary mass of the UHECRs}

 \author{Nicusor Arsene}
 \email[]{nicusorarsene@spacescience.ro}
 \affiliation{Physics Department, University of Bucharest, 
 Bucharest-Magurele, Romania}
 \affiliation{Institute of Space Science, P.O.Box MG-23, Ro 077125 
 Bucharest-Magurele, Romania}

 \author{Octavian Sima}
 \email[]{octavian.sima@partner.kit.edu}
 \affiliation{Physics Department, University of Bucharest, 
 Bucharest-Magurele, Romania}

 \author{Andreas Haungs}
 \email[]{andreas.haungs@kit.edu}
 \affiliation{Karlsruhe Institute of Technology, Institut f\"{u}r 
 Kernphysik, Karlsruhe, Germany}

 \author{Heinigerd Rebel}
 \email[]{Heinrich.Rebel@partner.kit.edu}
 \affiliation{Karlsruhe Institute of Technology, Institut f\"{u}r 
 Kernphysik, Karlsruhe, Germany}

 \date{\today}

\begin{abstract}
In this paper we study, using Monte Carlo simulations, the possibility to discriminate the mass of the Ultra High Energy Cosmic Rays (UHECRs) by 
combining information obtained from the maximum $X_{max}^{\mu}$ of the muon production rate  longitudinal profile of 
Extensive Air Showers (EAS) and the number of muons, $N^{\mu}$, which hit an array 
of detectors located in the horizontal plane. We investigate the 
sensitivity of the 2D distribution $X_{max}^{\mu}$ versus $N^{\mu}$ to the mass of the primary particle generating the air shower. 
To this purpose  we analyze a set of CORSIKA showers induced by protons and iron nuclei at energies of $10^{19}$eV and $10^{20}$eV,
at five angles of incidence, $0^{\circ}$, $37^{\circ}$, $48^{\circ}$, $55^{\circ}$ and $60^{\circ}$. 
Using the simulations we obtain the 2D Probability 
Functions $Prob(X_{max}^{\mu},N^{\mu} \ | \ p)$ and 
$Prob(X_{max}^{\mu},N^{\mu} \ | \ Fe)$ which give the probability that a shower induced by a proton or iron nucleus contributes to a specific point on the plane ($X_{max}^{\mu}$, $N^{\mu}$). Then we construct
the probability functions $Prob(p\ | \  X_{max}^{\mu},N^{\mu})$ and $Prob(Fe \ | \  X_{max}^{\mu},N^{\mu})$ which give the probability that a certain point on the plane ($X_{max}^{\mu}$, $N^{\mu}$) corresponds to a shower initiated by a proton or 
an iron nucleus, respectively. 
Finally, a test of this procedure using a Bayesian approach, confirms an improved accuracy of the primary mass estimation in comparison with the results obtained using only the $X_{max}^{\mu}$ distributions.
\end{abstract}

\maketitle

\section{Introduction}

The mass composition of the primary UHECRs together with their energy spectrum and arrival directions are the fundamental 
data
when searching for the sources and the acceleration mechanisms of the cosmic rays. 
Various detection techniques, such as surface detectors (scintillation modules \cite{::2013dga} or water Cherenkov tanks \cite{Ave:2007zz}), fluorescence detectors \cite{Abraham:2009pm}, \cite{Tokuno:2012mi}, radio antennas \cite{Abreu:2012pi}, microwave detection \cite{PhysRevLett.113.221101}, have been proposed to study these observables. 
Despite concerted efforts in many experiments,
such as Pierre Auger Observatory \cite{ThePierreAuger:2015rma}, Telescope Array \cite{AbuZayyad201287}, HiRES \cite{Sokolsky201174}, AGASA \cite{Chiba:1991nf}
to answer these fundamental questions, a clear answer is not yet given.  

In the present work we focus on the problem of the properties of the primary particle which initiates the EAS 
using the informations from the ground particle detectors.

One observable which is sensitive to the mass of the primary particle is the atmospheric depth where 
the density of the secondary charged particles reaches its maximum. 
This observable decreases roughly proportionally with the logarithm of the mass $A$ of the primary particle. Its sensitivity to $A$
is illustrated by the difference in the values for $p$ and $Fe$ induced showers of about $100$ g \, cm$^{-2}$ \cite{Aab:2015bza} at the same 
energy. It can be obtained experimentally by measuring the shower UV light with fluorescence detectors (FD) \cite{ThePierreAuger:2015rma},  \cite{Abraham:2009pm}, \cite{AbuZayyad201287}, \cite{Sokolsky201174}. Indeed, the intensity of UV light 
emitted from an elementary volume consequent to the excitation of the nitrogen molecules in the atmosphere by the secondary charged particles 
in EAS, is proportional with the charge density. Thus, with the FDs the dependence of the charged particle density on atmospheric depth 
can be obtained. The drawback of this technique is the low duty cycle of FD measurements (up to $\sim 15\%$ \cite{ThePierreAuger:2015rma}), 
due to the fact that the UV light from an EAS can be measured only during moonless nights and only in good atmospheric conditions. This fact, 
combined with the low statistics of the UHECRs at $E > 10^{19}$ eV, has a significant contribution to the uncertainty of mass reconstruction 
by FDs measurements.

To increase the observational duty cycle, 
the reconstruction of the primary mass on the basis of the signal of 
the surface detectors (duty cycle $\sim 100 \%$) would be advantageous.
This can be done 
using the reconstructed profile of the muon production depth (MPD) from EAS on the basis of the signal of 
the surface detectors, 
as proposed by Cazon et al. \cite{Cazon:2004zx, Cazon:2003ar} in the case of the Pierre Auger Observatory.
The individual muon production depth (the muon production point expressed in units of atmospheric depth) can be calculated using the muon arrival time in the detectors and the arrival time of the shower core.
Then, the longitudinal profile of the muon production rate can be obtained as the depth dependence of the number of muons produced per unit 
of atmospheric depth. The maximum $X_{max}^{\mu}$ of this profile was proposed as an observable sensitive to the primary mass.

The number of muons in the shower is also sensitive to the primary mass. However, it has a stronger dependence on 
the energy of the primary particle 
than on the primary mass, and due to this fact the uncertainty of energy determination has a high impact on mass discrimination 
using this observable.

In a preliminary study \cite{Arsene,Arsene_Sima}
we have shown that by using the information included in the correlation $X_{max}^{\mu}$ versus $N^{\mu}$, 
the accuracy of the primary mass reconstruction can be improved in comparison with the method which uses 
only the $X_{max}^{\mu}$ distribution. This correlation could also be used to test the high energy interaction 
models. Our preliminary study was based on simulations done with the CORSIKA code \cite{corsika,corsika1} using the 
thinning option, without applying a resampling scheme. In the present work the study is extended by applying 
the resampling scheme proposed by Billoir \cite{Billoir:2008zz}.
Also, the parametrization of the 2D distribution $X_{max}^{\mu}$ versus $N^{\mu}$ is improved. The study is 
done both in the case when $N^{\mu}$ corresponds to all the muons from a given radial range where the muon production depth is 
reconstructed from the arrival times of all these muons and in the realistic case when $N^{\mu}$ and the production depth correspond to 
the muons which hit the detectors from an array like  
AMIGA surface detector array   \cite{Wainberg:2013koa}, \cite{Videla:2015xia}, \cite{Aab:JINST2016}
of the Pierre Auger Observatory. 
In order to test the principle of the method, in this exploratory work the experimental uncertainties are not included and the 
detector simulation is not done. However, some results of the effects of uncertainties in the arrival time and in the reconstruction of 
the shower parameters are presented.

In Section II the observables $X_{max}^{\mu}$ and $N^{\mu}$ are introduced. 
In Section III the simulations used are presented 
and the data analysis for obtaining the muon production depth and the muon number is discussed; the resampling scheme applied 
is briefly described. In Section IV the 2D distribution $X_{max}^{\mu}$ versus $N^{\mu}$ is presented and parameterized. 
In Section V a Bayesian approach is applied in order to test the mass discrimination performance on the basis of this 2D distribution. 
Section VI concludes the paper.

\section{The $X_{max}^{\mu}$ and $N^{\mu}$ observables}

During the development of an EAS, various types of secondary particles are produced, which further interact in the atmosphere or decay.
Thus, the number of secondary particles increases after the first interaction, reaching a maximum 
at a certain atmospheric depth, where the value depends on the mass and energy of the primary particle.
The dependence of the number of charged particles on the atmospheric depth represents the longitudinal shower profile. 
The number of muons in the shower reaches a maximum 
on its development much deeper 
than the electromagnetic component, due to the increased production of muons when the energy of the parent pions decreases and 
to the larger mean free path of the muons in the atmosphere. 
Both the maximum of the charged particles longitudinal profile and the maximum of the longitudinal profile of the muon production 
rate are sensitive to the mass of the primary particle and
can also provide additional information 
useful to constrain the high-energy interaction models \cite{Aab:2014aea,Collaboration:2012wt}. 

$X_{max}^{\mu}$ can be evaluated after the reconstruction of the MPD. 
Experimentally the MPD can be reconstructed more accurately from the signal of the detectors from a specific radial range. 
This is due to the fact that the electromagnetic component of the shower can contribute to some extent to the signal of the muon detectors. Therefore, 
the detectors located close to the shower core, where the electromagnetic component has a much higher contribution, would introduce an 
uncertainty in the muon reconstruction. On the other hand, far from the shower core the number of muons decreases dramatically and also 
the uncertainty of the reconstruction of the MPD increases. Therefore, even if we do not simulate the detectors
in our analysis we reconstruct the MPD using the muon arrival time in the observational 
plane (the ground plane where the detectors are located) in several radial ranges, from 1000, 1400 or 1800 m to 4000 m. 

It is intuitive that for the 
same geometry of the shower axis, the mean number of muons on the ground will be higher for an iron induced shower compared to a proton shower of 
the same energy, due to the higher multiplicity at the first interactions.
In fact a gross estimation of the dependence of the number of muons on primary mass and energy 
can be obtained using the Matthews-Heitler model \cite{Matthews:2005sd}

\begin{equation}\label{eq:nmu_heitler}
N_{t}^{\mu} = A \left(\frac{E/A}{\xi_{\mathrm c}}\right)^\beta,
\end{equation}
where below the critical energy $\xi_{\mathrm c}$ all the charged pions are assumed to decay yielding muons, 
and the parameter $\beta$ $\simeq$ 0.9. As can be seen from this equation, $N_{t}^{\mu}$ has a strong, almost linear, dependence on energy whereas the dependence on mass 
is much weaker. Therefore, the direct use of $N_{t}^{\mu}$ for mass discrimination requires a very accurate determination of energy;
also, the evaluation of $N_{t}^{\mu}$ from the signal of the detectors requires a good description of the muon lateral distribution 
function, i.e. a good reproduction of the experimental dependence by the theoretical functions.

In our study the muon number is obtained from simulations and $N^{\mu}$ represents the number of muons which hit the detectors 
from a specific array with an energy threshold of 300 MeV. In addition, for the purpose of comparison with an ideal situation, we consider also the case when $N^{\mu}$ represents the total number of muons which reach the ground 
in a given radial range.

\section{Simulation data and evaluation of $X_{max}^{\mu}$ and $N^{\mu}$}
\subsection{Simulations}
The statistics of this analysis is based on $120$ CORSIKA simulations for each primary particle type ($proton$ and $iron$), energy ($10^{19}$ eV and $10^{20}$ eV) and incidence angle ($0^{\circ}$, 
$37^{\circ}$, $48^{\circ}$, $55^{\circ}$ and $60^{\circ}$). Thus in total 2400 simulated showers were analyzed. In the simulations the EPOS hadronic interaction model for high energies \cite{Pierog:2013ria} and FLUKA for low energies \cite{Ferrari:898301} were used. 
The thinning level (see Section C) was set to $10^{-6}$ and the maximum weight to 1000. For concreteness, 
the simulations were done with the Earth's magnetic field corresponding to the location of the Pierre Auger Observatory 
and the data analysis was based on a detector array with detector separation of 750 m, similar with the 
AMIGA array  \cite{Wainberg:2013koa}.

\subsection{Muon arrival times from EAS}

The idea of using the information of the muon arrival times in order to estimate the nature of the primary UHECR was previously 
studied in \cite{Brancus,Haeusler,Rebel:1994ed} in the context of the KASCADE experiment \cite{Antoni:2003gd} and later in 
\cite{Cazon:2004zx,Cazon:2003ar,Andringa:2011ik,Aab:2014dua}. The principle of the method is to reconstruct the longitudinal 
distribution of the MPD in EAS based on the times when the shower muons reach
the ground relative to the time when the shower core reaches the ground. The method is applicable to experiments which can record 
the temporal signal of the secondary particles at ground level, such as the Pierre Auger Observatory. 
One of the advantages of this method is due to the duty cycle of the surface detectors which is $\sim 100 \%$, and therefore 
much higher than of the fluorescence detectors.  

\begin{figure}[t]
\centering
\includegraphics[scale=0.3]{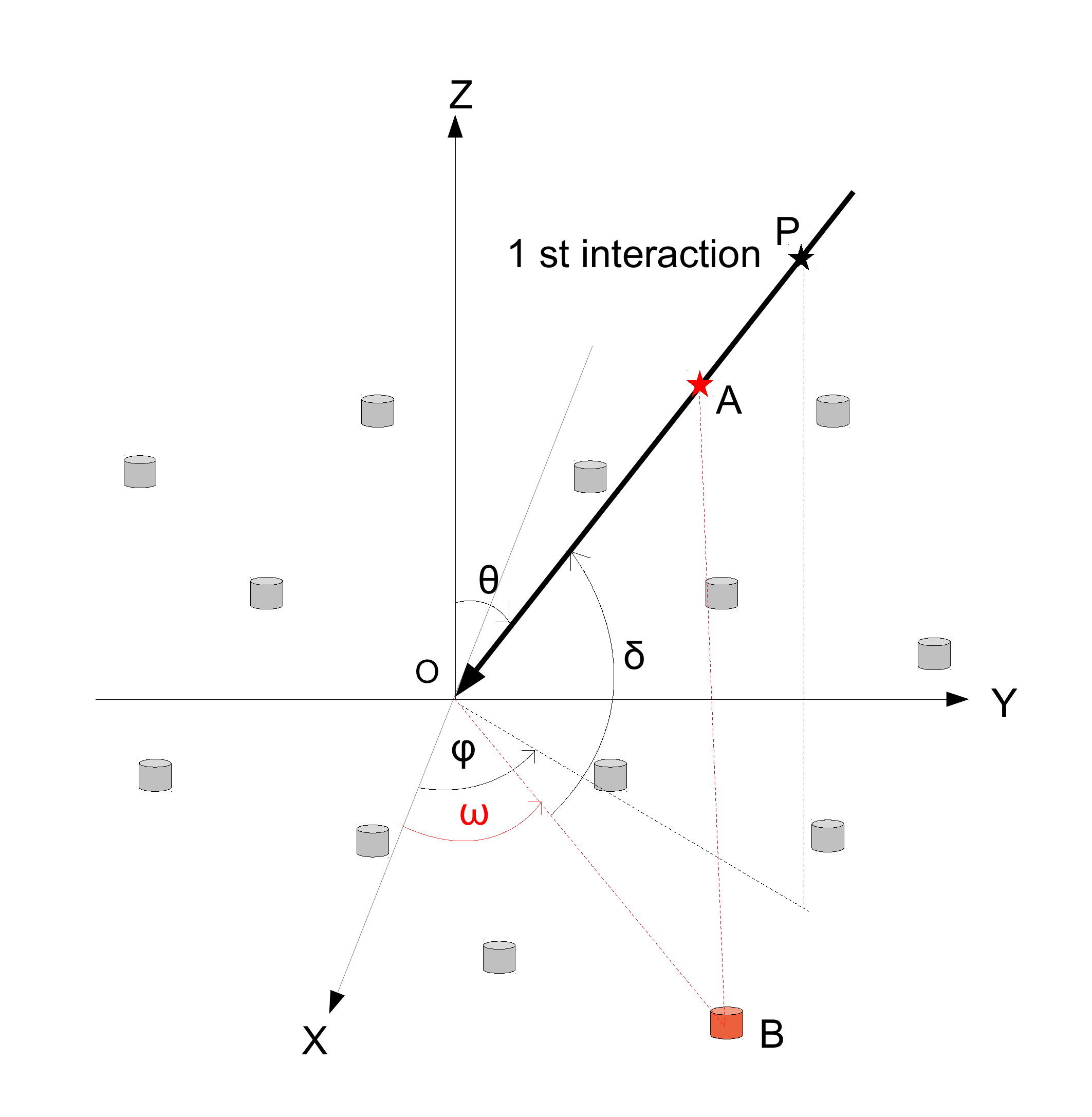}
\caption{Coordinate system of the EAS according with CORSIKA \cite{Arsene_Sima}. $P$ = point of the first interaction, $A$ = muon production point, $O$ = shower core on ground, $B$ = location of the muon detector.}
\label{coordinates}
\end{figure}

The lifetime of the muons in the EAS is quite large 
and the deviation in the Earth's magnetic field is very small, so one can consider that 
the muons travel in straight lines through the atmosphere from the production point, close to the shower axis, to the ground.
Thus, if the kinematic delay \cite{Cazon:2003ar} and the scattering effects are small, the muon production locus can be calculated 
using the difference between the time $t_{\mu}$ of 
arrival of muons in the detector and $t_c$ of the shower core at ground. The basic idea is the following. Consider a shower in 
which the first interaction of the primary UHECR takes place in the point $P$ (Fig. \ref{coordinates}). A muon is produced in 
the point $A$ at time $t_0$ 
(the arrival time of the core in $A$)
and registered at time $t_{\mu}$ in a detector located in $B$. The shower core reaches the ground in point $O$ at time $t_c$. 
The difference between the pathlengths $AB$ of the muon and $AO$ of the shower core is equal to $v_{\mu}\ (t_{\mu}-t_0)-c \ (t_c-t_0)$, 
where $v_{\mu}$ is the average muon speed and $c$ is the speed of the shower front (speed of light); if the kinematic correction is 
negligible, $v_{\mu}$ is practically equal also with $c$.

In simulated showers the distance $OB$ is known, as well as the angles $\theta$ and $\phi$ of the shower axis and $\omega$ of 
the direction towards the detector. Then, with known values of the lengths of $OB$ and $AB-AO$ and known angle in $O$, 
the triangle $AOB$ can be resolved and the muon production point $A$ can be determined. More precisely,

\be
AO=c(t_c-t_0) = \frac{OB^2 - c^2 (t_{\mu} -t_c)^2}{2 [c (t_{\mu} - t_c) + OB \cos \delta]}
\ee

where $\delta$ is the angle between the shower axis and the direction $OB$.

Using this equation the atmospheric depth of the muon production can be 
calculated.
Of course, we consider that the muon was produced on the shower core only if the properties of the triangle $AOB$ are fulfilled 
(if the difference in the muon arrival time and the arrival time of
the shower core is compatible with the difference in the traveled paths \cite{Arsene}). Note that since the MPD is reconstructed using the coordinates of the muons in the observation plane, 
not in the shower plane (perpendicular to the shower axis), the problems arising from the projection of the particle coordinates from 
observation plane to the normal plane (see \cite{Sima:2011zz}) are avoided.

After the reconstruction of the longitudinal muon profile, the maximum of the distribution $X_{max}^{\mu}$ 
 is obtained by fitting the profile with the Gaisser-Hillas function \cite{GH}.

In order to estimate the uncertainty of $X_{max}^{\mu}$ reconstruction due to the uncertainty of the arrival times and of the 
 reconstruction of the position of the shower core, we analyzed a sample of 140 sets of simulated data. Each set was obtained from the same parent CORSIKA output file by applying a Gaussian spread with $\sigma_{t_{\mu}}$ = 20 ns to each arrival time and with 
 $\sigma_{x} = \sigma_{y}$ = 50 m to the position of the shower core. The parent shower was induced by a proton at $E = 10^{19}$ eV, $\theta = 0^{\circ}$ and we considered all the 
 muons from the radial range R = [1800 - 4000 m]. The standard deviation of the $X_{max}^{\mu}$ for this set of data was 1.9 g/cm$^2$.

\subsection{Resampling}
The simulation process of the EAS is extremely time consuming and it requires a very large amount of data storage for the 
cascades induced at energies which exceed $\sim 10^{17}$eV. The thinning method is implemented in the CORSIKA \cite{corsika1} code 
for reducing the computation time and the output size by replacing, in certain conditions, a bunch of secondary particles by 
a single representative particle with a weight equal to the sum of the weights of the replaced particles. The cost of this procedure 
is that large, artificial, uncertainties will be introduced. 
An extreme example of artificial fluctuations one can imagine when several detectors are placed in a radial range were in average actually one 
particle 
hits each detector. While in the simulation the particles have a typical weight of 100, according to simulations, only one detector 
from 100 detectors will be hit 
by a particle (with a weight of 100) and the others will have no hit. Thus, the spread from one detector to the other of the particle 
density reconstructed by the detectors will be much higher in the simulations than in the actual shower.
In order to reduce the fluctuations associated with the strong thinning scheme, we applied for all CORSIKA simulations the "resampling" 
procedure proposed in \cite{Billoir:2008zz}.

The procedure consists in regenerating the particles around the detectors according to a Poisson distribution in a \textit{sample region} around 
detectors. The area of this sample region $(A_{sr})$ depends on the distance to the shower axis and the nature of resampled particles (electrons, 
muons, hadrons) (for more details see \cite{Billoir:2008zz}). In our case we have chosen a radial dimension $\sim 0.02 \times r_{i}$, where $r_{i}$ 
is the distance of the detector $"i"$ to the shower axis in the observational plane of the cascade.
Instead of each particle with weight = $n$ a number of $n$ particles will be generated with the weights equal to 1 and the same nature, energy, arrival direction and times 
inside the sample region. Their positions are sampled from a 2D Gaussian distribution with $\sigma_{sr} = r_{sr}^{i}/2$, where $r_{sr}^{i}$ is the radius of the sample region.
The arrival times of the regenerated particles are then updated, including a smearing of the type $t^{'} = t \times exp(\sigma_{t} G)$ where $G$ is 
a random number from a Gaussian distribution centered on $0$ and variance 1, and $\sigma_{t} = 0.1$.

Due to the fact that a much higher number of particles represent the output of the resampling procedure, the fluctuations are much 
reduced. 

In order to evaluate $X_{max}^{\mu}$ and $N^{\mu}$ a first code was developed to obtain a file containing the results of the resampling
procedure applied to the data read from the original CORSIKA output files. Then the resampled files were processed for obtaining 
$X_{max}^{\mu}$ on the basis of arrival times and $N^{\mu}$
by counting the muons.

The data analysis was done with a code developed in the ROOT framework. Two analysis runs were done: in the first 
$X_{max}^{\mu}$ and $N^{\mu}$ were obtained by analyzing the information provided by all the muons that reach the ground in 
a given radial range (the \textit{ideal} case), whereas in the second run only the information pertaining to the muons that 
hit the detectors located in that range was used (the \textit{detector array} case).

\section{Results: $X_{max}^{\mu}$ vs. $N^{\mu}$ sensitivity to the primary mass}

It is obvious that at the same energy, on average a lighter particle will travel more deeply in the atmosphere before the first 
interaction 
than a heavier nucleus and therefore the induced EAS 
will have a larger $X_{max}^{\mu}$ value. Thus $X_{max}^{\mu}$ is a mass sensitive observable. At the same time,
the multiplicity of the secondary particles from EAS at a certain energy depends on the mass of the primary particle, 
and thus the number of muons detected in the observational plane is another mass sensitive observable. 
It seems plausible that the correlated use of  
the information included in the maximum of the
longitudinal profile of the muon production rate
$X_{max}^{\mu}$ and in  the number of muons $N^{\mu}$ may provide improved accuracy of the primary mass reconstruction. 
Specifically, showers induced by protons are expected to have a more important contribution in some regions of the 2D distribution 
$(X_{max}^{\mu},N^{\mu})$, while showers induced by iron nuclei, 
in other regions, and thus the correlation $(X_{max}^{\mu},N^{\mu})$ should provide better mass discrimination than each independent
observable.Our aim is to test this conjecture.

For the mass reconstruction the Probability Functions
$Prob(p\ | \ X_{max}^{\mu},N^{\mu})$ and $Prob(Fe\ | \ X_{max}^{\mu},N^{\mu})$ are required. 
These functions give the probability that a certain point from the plane
$(X_{max}^{\mu}$, $N^{\mu})$ 
corresponds to a shower induced by a proton or an iron nucleus, respectively.
These functions were obtained as follows. 

First, the 2D distribution $X_{max}^{\mu}$ versus $N^{\mu}$ was constructed for each primary particle, energy and incidence angle using 
the simulations. 
In Figures \ref{e20_N_X_R_1800_} and \ref{e20_N_X_R_1800_infill_} we represented this distribution obtained by analyzing all the 
showers with $E=10^{20}$ eV and different zenith angles.
In Figure \ref{e20_N_X_R_1800_} (\textit{ideal} case) the reconstruction of the muon production depths was done using all the muons
which reach the observational plane in the radial range [1800, 4000 m], whereas in Figure \ref{e20_N_X_R_1800_infill_} 
(\textit{detector array} case) only the muons which hit the 
detectors located in the same radial range from the considered array
were analyzed. In the latter case, significantly reduced information is available, because a very small percentage of the total 
number of muons reach the detectors, making it more complicated to distinguish between the species of the 
primary particle which induced the shower. 
In order to increase the statistics in the 
detector array case, each CORSIKA shower was used 10 times, by randomly distributing the shower core over the array.

Next, 1D distributions of $N^{\mu}$ and $X_{max}^{\mu}$ obtained by projecting the 2D distributions were analyzed to test whether a simple analytical parametrization could be used. In Figures  \ref{1D_N} and \ref{1D_X} these distributions are displayed for showers with $E=10^{20}$ eV and $\theta=37^{\circ}, 48^{\circ}$, and 55$^{\circ}$, in the detector array case. The $N^{\mu}$ distributions 
resemble
Gaussian functions. Concerning the distribution of the values of $X_{max}^{\mu}$, it is important to observe that the information on the MPD obtained from the detectors located in a specific radial range comes mostly from a specific range of atmospheric depths. For example, at high angles the muons generated at high atmospheric depth contribute less to the reconstruction of the distribution of the production depth. Thus the observed distribution appears truncated with respect to
the true MPD distribution.
 This can be seen in Figure  \ref{1D_X} for proton showers at  $\theta=55^{o}$. 
The shape of $X_{max}^{\mu}$  can be approximated by a Gaussian distribution within the range of reconstructed values, so that the observed distribution can be described by a truncated Gaussian function.
In the same case, the distribution for iron showers does not present this feature, and there are no values in the iron case beyond the cut observed in the proton distribution. 
Thus, in view of the final goal of identifying the primary particle on the basis of the distributions, 
as a first approximation,
the truncation in the Gaussian functions can be neglected, 
because in the range beyond the cuts there are no $X_{max}^{\mu}$ values which could be wrongly attributed due to the 
truncation being neglected.

Using these results the probability density functions
$Prob (X_{max}^{\mu},N^{\mu} \ | \ p)$ and $
Prob (X_{max}^{\mu},N^{\mu} \ | \ Fe)$ were evaluated and parameterized by:
\begin{widetext}
\begin{eqnarray}
& & Prob (X_{max}^{\mu},N^{\mu} \ | \ cr)= \,
\frac {1}{2\pi \sigma_{X_{max}^{\mu}} \sigma_{N^{\mu}} \sqrt{1-\rho^2}}\, \times \nonumber \\ & \times & \exp \!\! \left[- \frac{1}{2 (1-\rho^2)}\!\!
\left[\!\! \left( \frac{X_{max}^{\mu}-<\!\!X_{max}^{\mu}\!\!>}{ \sigma_{X_{max}^{\mu}}} \right)^2 \right. \right. \!\! + \!\! \left. \left.
\left( \frac{N^{\mu}-<\!\!N^{\mu}\!\!>}{ \sigma_{N^{\mu}}} \right)^{2}\!\! -\!\! 2 \rho \!\left( \frac{X_{max}^{\mu}-<\!\!X_{max}^{\mu}\!\!>}{ \sigma_{X_{max}^{\mu}}} \right)\!\!
\left( \frac{N^{\mu}-<\!\!N^{\mu}\!\!>}{ \sigma_{N^{\mu}}}\right)\!\! \right] \right]
\label{2DProbFunc}
\end{eqnarray} 
\end{widetext}
which is a two dimensional Gaussian function of variables $X_{max}^{\mu}$ and $N^{\mu}$; here $cr$ represents the primary cosmic ray, i.e. $p$ or $Fe$.
We mention that in the preliminary study \cite{Arsene_Sima} the correlation between $X_{max}^{\mu}$ and $N^{\mu}$ was neglected in equation \ref{2DProbFunc}.

\begin{figure}
  \includegraphics[height=.14\textheight ]{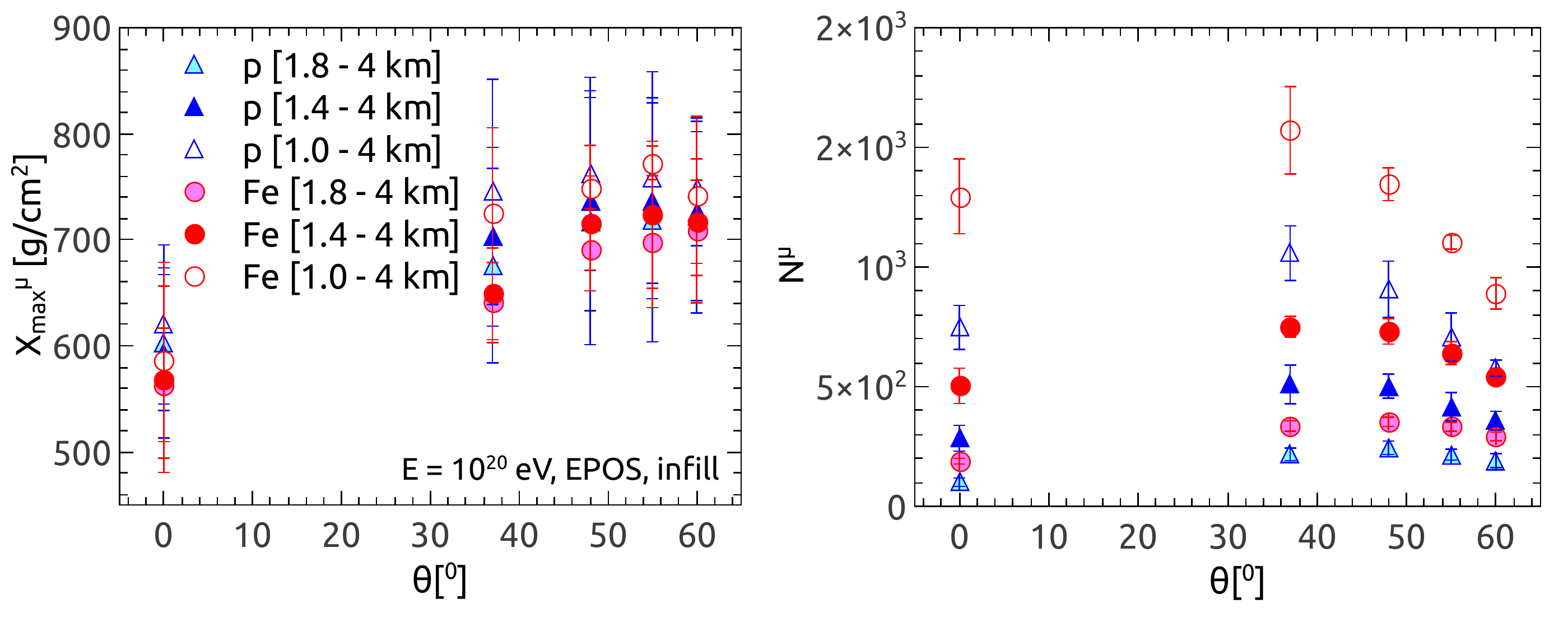}
  \caption{The parameters obtained by fitting the 2D distributions with Eq. \ref{2DProbFunc} for the showers 
  induced at $E = 10^{20}$eV. The plots illustrate the 
  dependence of the parameters $X_{max}^{\mu}$ and $N^{\mu}$ on the radial range at the ground level and the dependence on the zenith angle of the shower axis. In each case the information provided by all the muons 
 \textit{(top)}, or only by the muons which hit the  detectors \textit{(bottom)} is used for $X_{max}^{\mu}$ and $N^{\mu}$ evaluation.}
  \label{fit_e20_infill}
\end{figure}

One can observe that the $X_{max}^{\mu}$ values increase when the zenith angle of the shower axis increases. The evolution of the parameters
$<X_{max}^{\mu}>$ and $<N^{\mu}>$ obtained by fitting the 2D distributions with the function from equation \ref{2DProbFunc} for different zenith angles and 
different radial intervals in observational plane is shown in Figure \ref{fit_e20_infill}. Of course, if we consider a larger radial range in the observational 
plane, the number of muons which contribute to the longitudinal distribution will increase and the $X_{max}^{\mu}$ value will be obtained 
with smaller uncertainty,
but in view of the discussion from Section II we restricted the analysis to detectors located in defined radial ranges. The data presented in Figures  \ref{e20_N_X_R_1800_} and \ref{e20_N_X_R_1800_infill_} and analyzed further correspond to the radial range from 1800 to 4000 m.

\begin{widetext}

\begin{figure}
\centering
\includegraphics[height=.35\textheight]{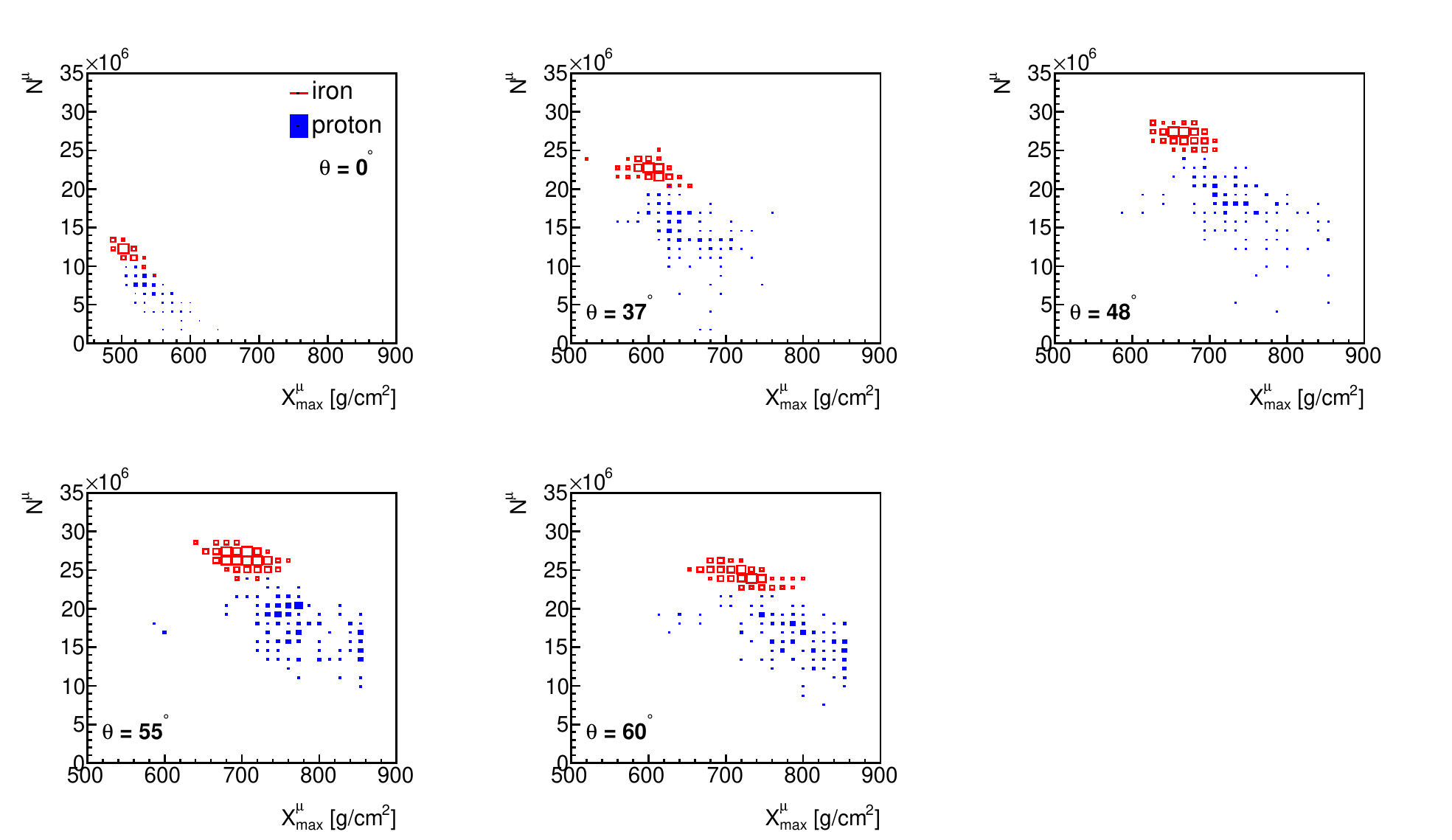}
\caption{Event by event analysis, number of muons at ground level in the radial range [1800 - 4000 m]  versus $X_{max}^{\mu}$. All the muons from this radial range were analyzed for obtaining $X_{max}^{\mu}$ and $N^{\mu}$.
120 CORSIKA simulations per case, at $E=10^{20}$eV and five zenith angles.}
\label{e20_N_X_R_1800_}
\end{figure}

\begin{figure}
\centering
  \includegraphics[height=.35\textheight]{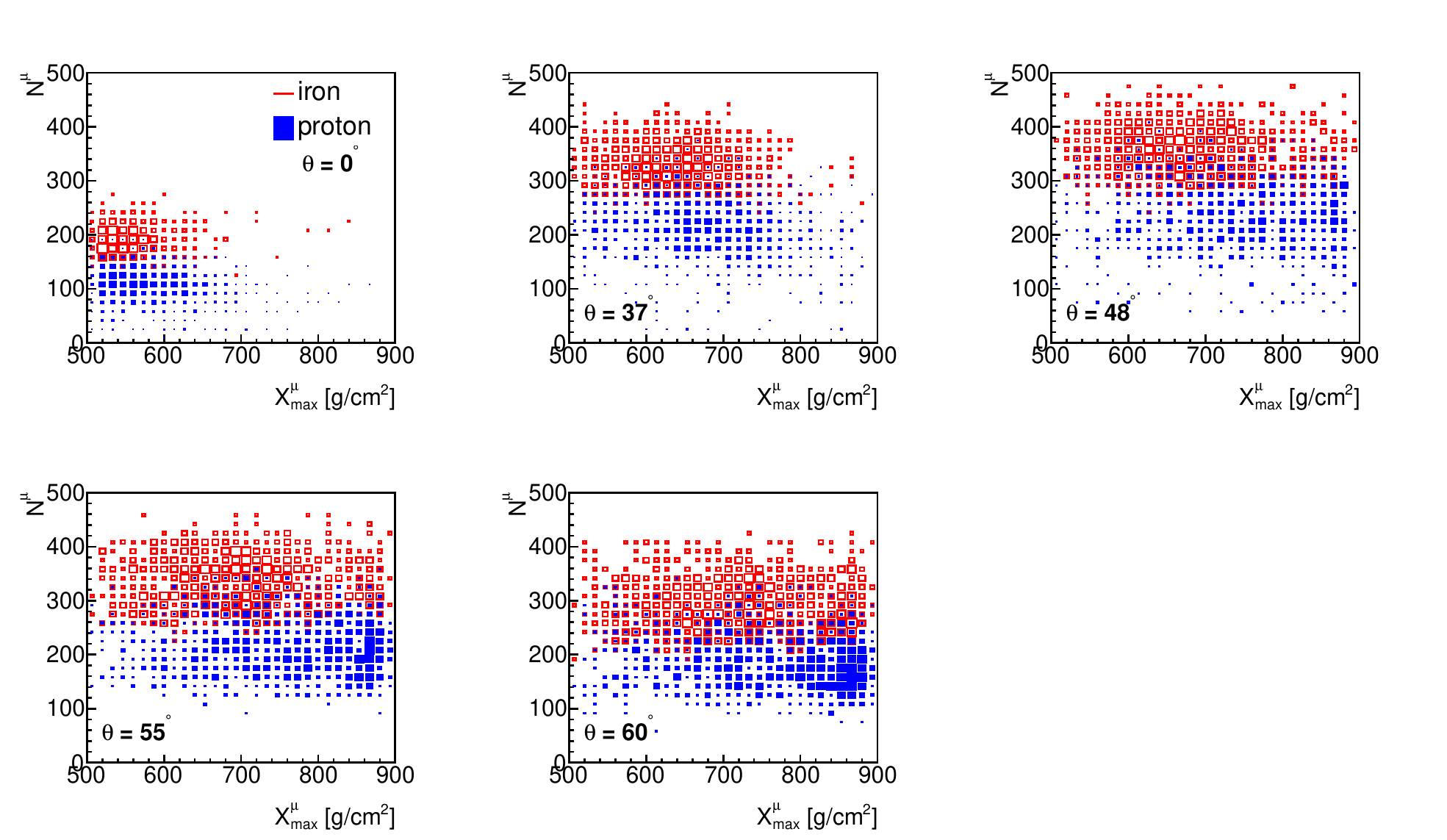}
  \caption{Event by event analysis, number of muons at ground level in the radial range [1800 - 4000 m] versus $X_{max}^{\mu}$. The information on $X_{max}^{\mu}$ and $N^{\mu}$ is obtained by analyzing the muons which hit the detectors from this radial range.
  1200 CORSIKA simulations per case, at $E=10^{20}$eV and five zenith angles.}
  \label{e20_N_X_R_1800_infill_}
\end{figure}

\end{widetext}

\begin{widetext}

\begin{figure}
\centering
\includegraphics[scale=.93]{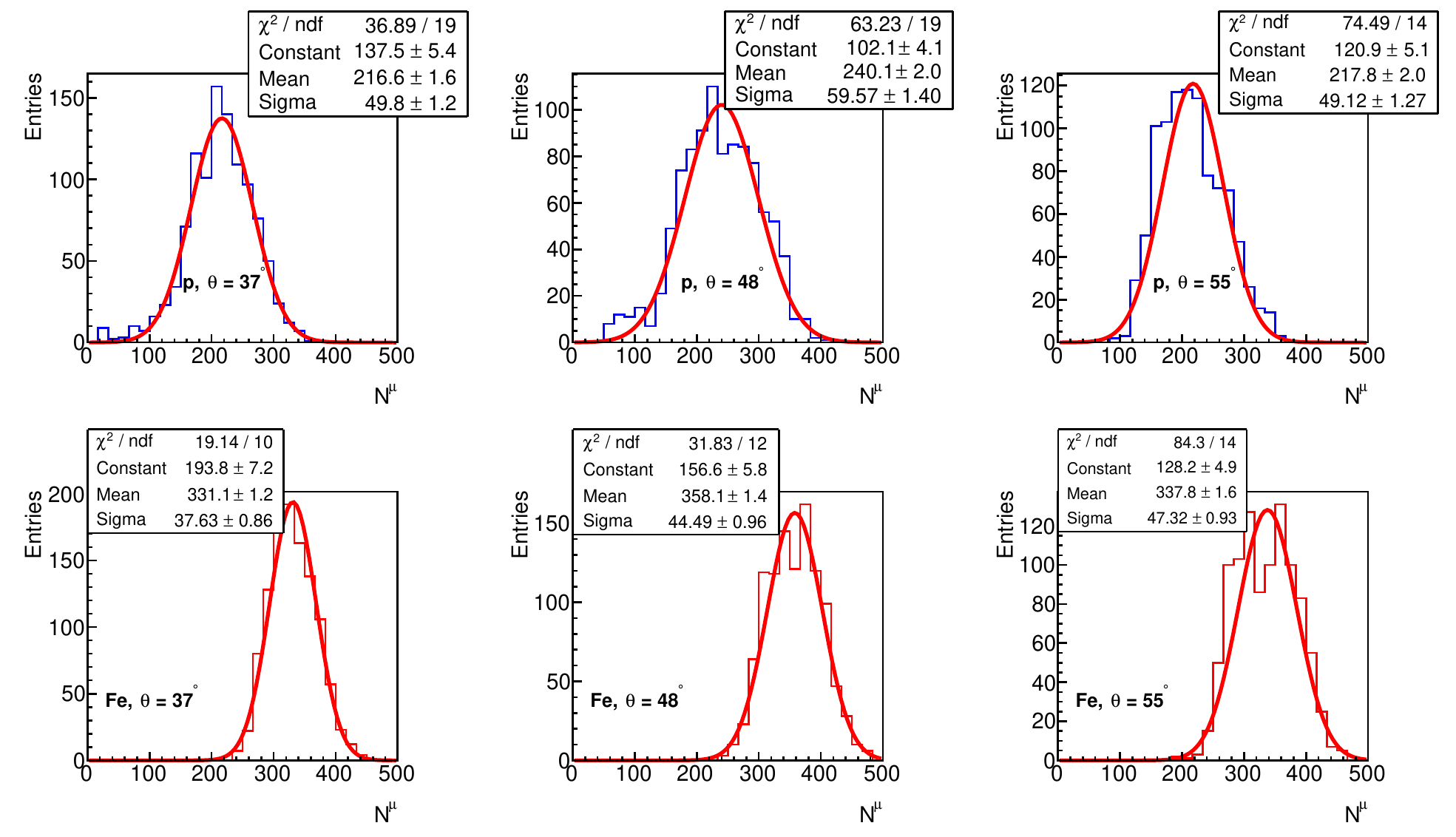}
\caption{Distribution of the number of muons at ground level in the radial range [1800 - 4000 m] in the detector case. Same conditions as in Figure \ref{e20_N_X_R_1800_infill_}.}
\label{1D_N}
\end{figure}

\begin{figure}
\centering
  \includegraphics[scale=0.93]{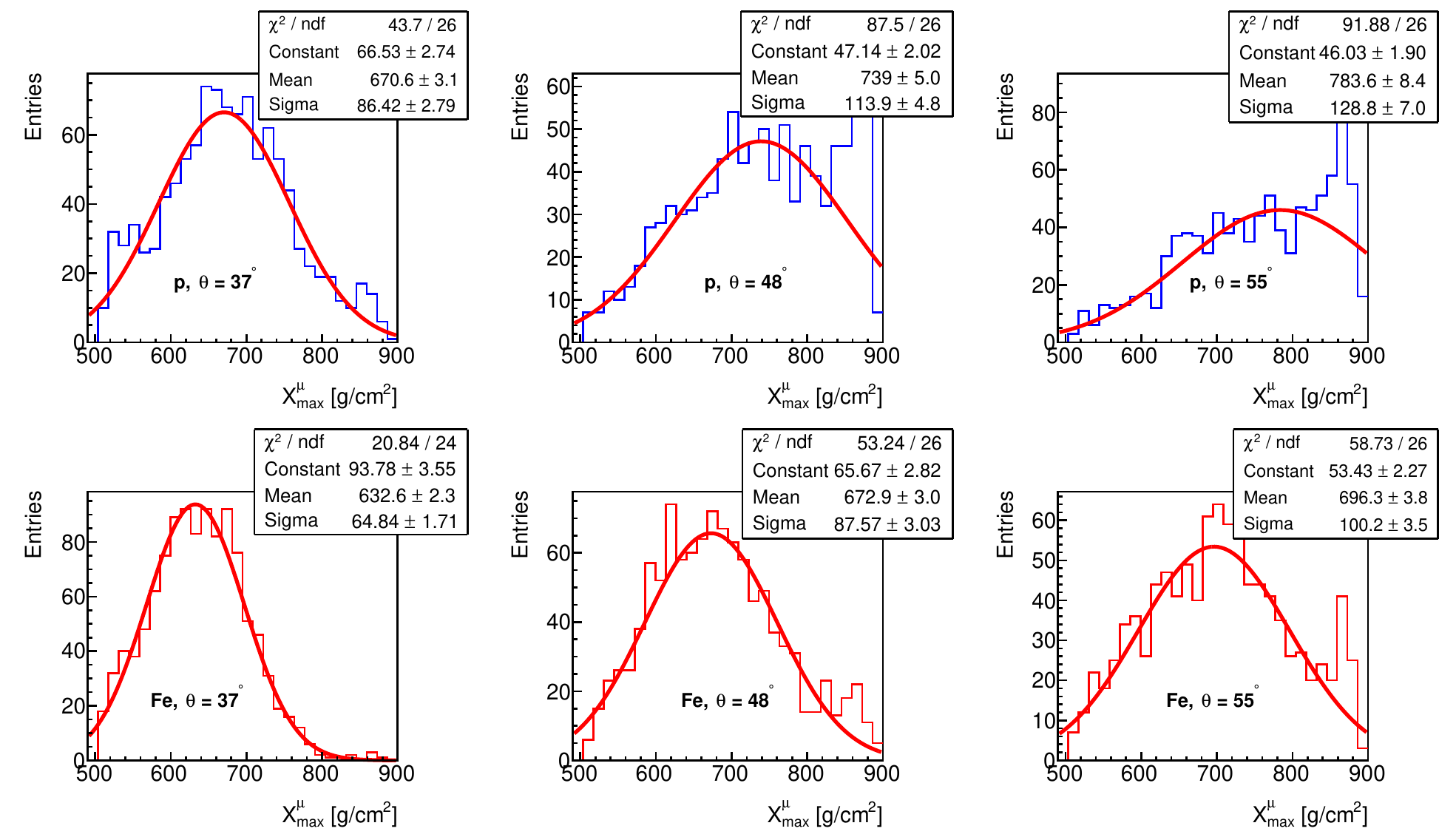}
  \caption{Distribution of  $X_{max}^{\mu}$ obtained from the arrival time information of muons detected in the  radial range [1800 - 4000 m].  Same conditions as in Figure \ref{e20_N_X_R_1800_infill_}.}
  \label{1D_X}
\end{figure}

\end{widetext}


\section{Bayesian test}

\begin{widetext}

\begin{figure}
  \includegraphics[height=.35\textheight ]{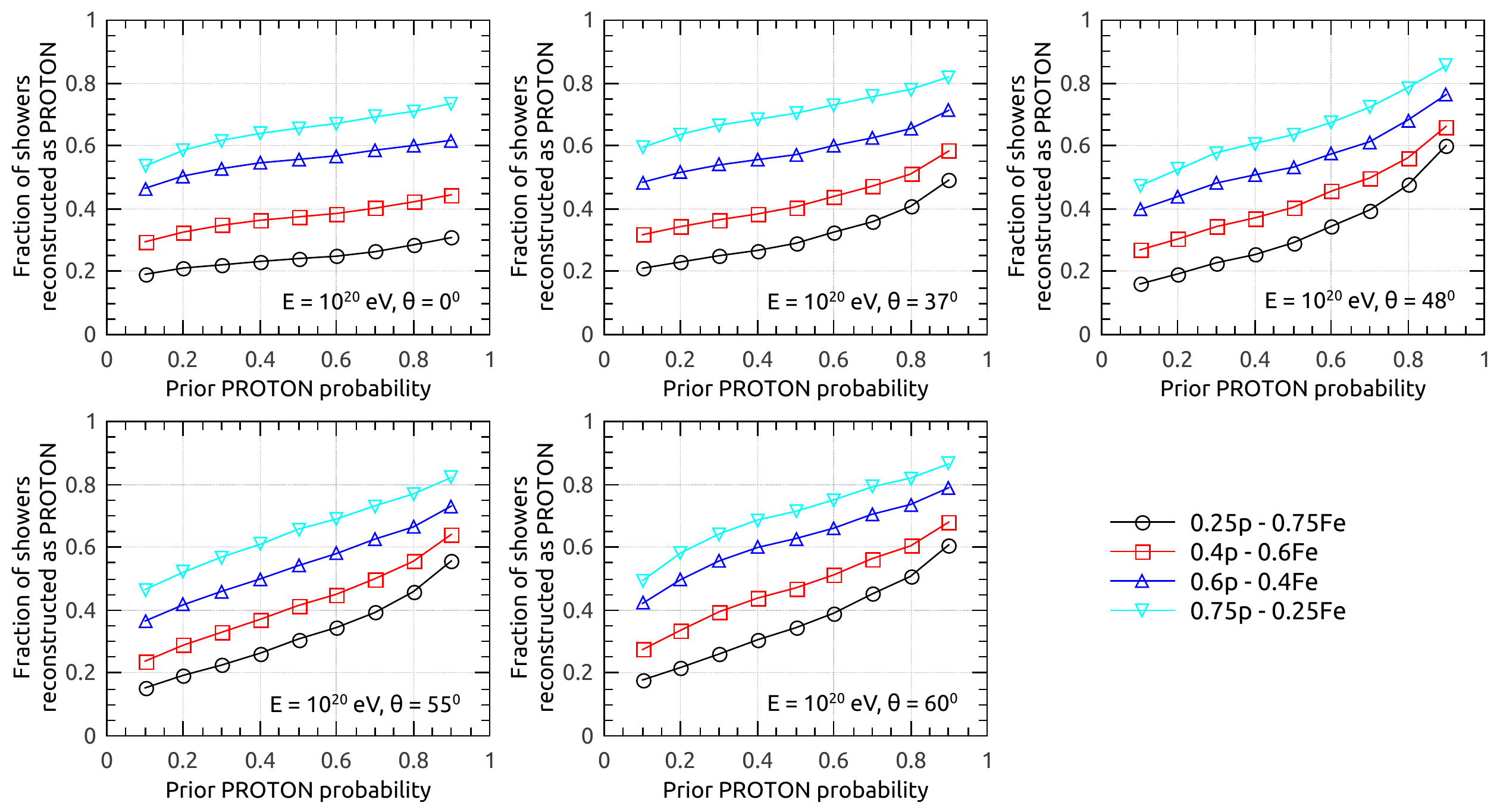}
  \caption{Potential mass discrimination of the method. Fraction of showers reconstructed as "PROTON" for different prior probabilities, different mixtures of showers and different zenith angles at $E = 10^{20}$eV (see text). Only the muons which hit the detectors are used for the analysis.}   
  \label{Bayes_p20}
\end{figure}

\end{widetext}

Using the Bayesian approach to test this procedure we need to define certain \textit{Prior} probabilities of proton and Fe showers and then calculate the 
\textit{Posterior} probabilities that a certain point from the plane $(X_{max}^{\mu}$, $N^{\mu})$ corresponds to a shower induced by 
a proton or an iron nucleus. $Prob(X_{max}^{\mu},N^{\mu} \ | \ p)$ and $Prob(X_{max}^{\mu},N^{\mu} \ | \ Fe)$ represent the probability 
to obtain the point with the coordinates $X_{max}^{\mu}$ and $N^{\mu}$ if the primary particle was a proton or an iron nucleus.
The posterior probability represents the probability that a point with the coordinates $(X_{max}^{\mu}, N^{\mu})$ is due to a shower 
initiated by a proton or by an iron nucleus. Supposing certain \textit{Prior} probabilities $Prob_{i}(p)$ and $Prob_{i}(Fe)$ which 
represent the abundance ratio of the primary protons and iron nuclei, we can calculate the \textit{Posterior} probabilities:
\begin{eqnarray}
  & & Prob_{a}(p \ | \ X_{max}^{\mu}, N^{\mu}) = \nonumber \\ &K& \cdot Prob(X_{max}^{\mu}, N^{\mu} \ | \ p) \cdot Prob_{i}(p) , 
\end{eqnarray}
\begin{eqnarray}
& & Prob_{a}(Fe \ | \ X_{max}^{\mu}, N^{\mu}) = \nonumber \\ &K& \cdot Prob(X_{max}^{\mu}, N^{\mu} \ | \ Fe) \cdot Prob_{i}(Fe) , 
\end{eqnarray}
where the constant $K$ can be calculated from the normalization: 
\begin{eqnarray}
\!\! \!\! \!\! \!\!Prob_{a}(p \ | \ X_{max}^{\mu}, N^{\mu}) + Prob_{a}(Fe \ | \ X_{max}^{\mu}, N^{\mu}) = 1.
\end{eqnarray}
To prove the stability of the method, we vary the prior probabilities between $0.1$ to $0.9$ and check the ability of the method to reconstruct the fraction 
of the showers close to the true values. In other words, the posterior probability should indicate the actual fraction of the showers from 
different mixtures. The results of this analysis are plotted in Figure \ref{Bayes_p20} for the showers with $E = 10^{20}$eV, 
thinning level $10^{-6}$ and different zenith angles.

\begin{figure}
  \includegraphics[height=.14\textheight ]{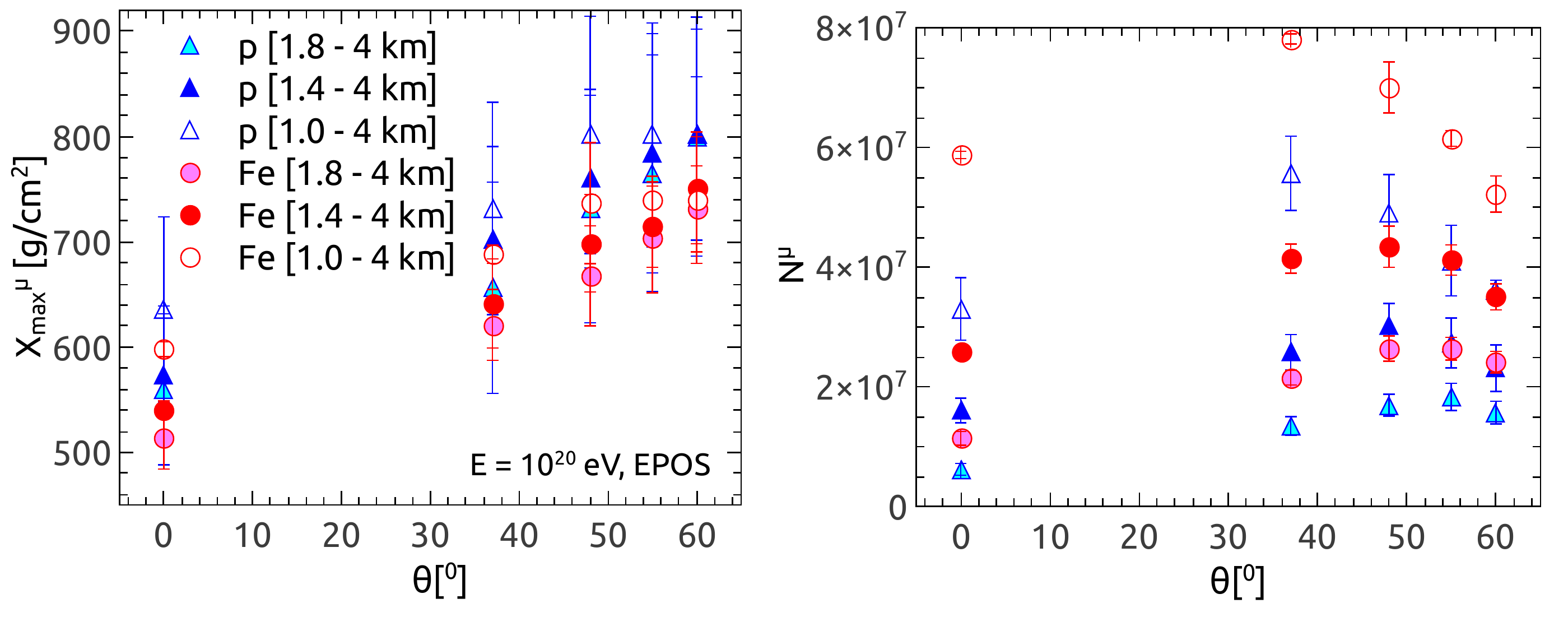}
  \label{fit_e20}
\end{figure}


We made a direct comparison to quantify if there is an improvement in the accuracy of the primary mass reconstruction using this method of 2D Probability Function 
against the method which uses only the $X_{max}^{\mu}$ distribution.
We applied these two methods for the simulations at $E = 10^{19}, 10^{20}$eV
and $\theta = 37^{\circ}$. We considered a mixture of 50\% proton and 50\% iron induced showers and the favorable case when the correct prior probabilities are used, i.e. the prior probabilities are $= 50\%$.
We find that the reconstruction accuracy increases to $\sim 98\%$ for the method which uses the two observables ($X_{max}^{\mu}$ versus $N^{\mu}$).
The results are listed in Table \ref{tabel}.

\begin{table}
\centering
\begin{tabular}{lllll}
\hline
                                & \multicolumn{2}{l}{\,\,$X_{max}^{\mu}$}  & \multicolumn{2}{l}{$X_{max}^{\mu} vs. N^{\mu}$}     \\
\hline
$E [eV]$                        & $10^{19}$ & $10^{20}$ & $10^{19}$ & $10^{20}$ \\
\hline
$Prob_{p \rightarrow p}[\%]$    & 40        & 44        & 94        & 97        \\
$Prob_{Fe \rightarrow Fe}[\%]$  & 52        & 56        & 96        & 98        \\
\hline
\end{tabular}
\caption{Mass reconstruction accuracy of the methods based on $X_{max}^{\mu}$ and on the 2D distribution ($X_{max}^{\mu}, N^{\mu}$). $Prob_{p \rightarrow p}$ represents the 
probability of correctly reconstructing the primary particle for a shower initiated by a proton (see text).}
\label{tabel}
\end{table}
We emphasize that the results obtained in this analysis do not include experimental uncertainties. The uncertainties 
of the values of the arrival times and of the reconstruction of the shower core position and axis angles will affect $X_{max}^{\mu}$ 
as presented in the end of Section III B. The uncertainty of the reconstructed energy of the shower will deteriorate the quality of 
information on mass of the primary particle embedded in the reconstructed value of $N^{\mu}$; in fact, this is the reason why the number
of muons is not directly used as a mass estimator  \cite{Aab:2014pza}. Indeed, if showers with the energy spread in a range instead
of showers with fixed energy are used to construct figures similar to Figures \ref{e20_N_X_R_1800_} and \ref{e20_N_X_R_1800_infill_}, 
then the separation between the $N^{\mu}$ distributions for $p$ and $Fe$ induced showers will deteriorate. For example, the width of 
the distributions will increase from ${\sigma}_{N^{\mu}} = 1.04 \times 10^{5}$
to ${\sigma}_{N^{\mu}} = 4.22 \times 10^{5}$ for $Fe$ induced showers
if instead of $E = 10^{19}$ eV and ${\theta}=48^{\circ}$ the energy and the incidence angle will be distributed in the ranges 
[$10^{18.9}, 10^{19.1}$] eV and [$46^{\circ}, 50^{\circ}$]. The width of $N^{\mu}$ distribution for proton induced showers,
which is rather large already for fixed proton energy, is practically insensitive to the relatively narrow distribution of primary energy.
The difference of the average values of $N^{\mu}$ between proton and iron showers is practically the same in the case of fixed energy
and angle
as in the case of distributed values.
Thus, in the presence of experimental uncertainties the mass discrimination 
power of the correlation ($X_{max}^{\mu}$,  $N^{\mu}$) will be lower than that obtained in this work, but it will still remain higher 
than in the case when only $X_{max}^{\mu}$ is used for mass discrimination.

\vspace{2.0em}

\vspace{2.0em}

\section{Conclusions and Outlook}

In this work, using CORSIKA simulations, we evaluated the possibility of discriminating the mass of the primary cosmic rays on the 
basis of the MPD 
taking into account the 2D distributions $X_{max}^{\mu}$ versus $N^{\mu}$. 
Because both $N^{\mu}$ and $X_{max}^{\mu}$ depend on the mass of the primary particle, but in a different way, 
the 2D distribution may contain more information on the mass of the primary cosmic ray than the individual distributions. 
Using this distribution we constructed the Probability Functions $Prob(p \ | \  X_{max}^{\mu},N^{\mu})$ and $Prob(Fe \ | \  X_{max}^{\mu},N^{\mu})$ 
which give the probability that a certain point from the plane $(X_{max}^{\mu}$ , $N^{\mu})$ corresponds to a proton or an iron shower.
We qualitatively found that the mass reconstruction accuracy improves when the  information from the correlation  $X_{max}^{\mu}$ 
versus $N^{\mu}$ is used in comparison with the method based only on the $X_{max}^{\mu}$ distribution.


\subsection*{Acknowledgments}
The authors 
are grateful
for the support offered by the colleagues from KIT and Pierre Auger Collaboration, especially Dr. R. Engel, D.M. J. Oehlschlaeger and Dr. D. Veberi\u{c}. O. Sima acknowledges support from the Romanian Authority for Scientific Research ANCS UEFISCDI project nr. 194/2012.

\bibliography{MPD_Nmu_v3_OS}

\begin{thebibliography}{35}
\expandafter\ifx\csname natexlab\endcsname\relax\def\natexlab#1{#1}\fi
\expandafter\ifx\csname bibnamefont\endcsname\relax
  \def\bibnamefont#1{#1}\fi
\expandafter\ifx\csname bibfnamefont\endcsname\relax
  \def\bibfnamefont#1{#1}\fi
\expandafter\ifx\csname citenamefont\endcsname\relax
  \def\citenamefont#1{#1}\fi
\expandafter\ifx\csname url\endcsname\relax
  \def\url#1{\texttt{#1}}\fi
\expandafter\ifx\csname urlprefix\endcsname\relax\def\urlprefix{URL }\fi
\providecommand{\bibinfo}[2]{#2}
\providecommand{\eprint}[2][]{\url{#2}}

\bibitem[{\citenamefont{Apel et~al.}(2013)}]{::2013dga}
\bibinfo{author}{\bibfnamefont{W.~D.} \bibnamefont{Apel}} \bibnamefont{et~al.},
  \bibinfo{journal}{Astropart. Phys.} \textbf{\bibinfo{volume}{47}},
  \bibinfo{pages}{54} (\bibinfo{year}{2013}).

\bibitem[{\citenamefont{Ave et~al.}(2007)}]{Ave:2007zz}
\bibinfo{author}{\bibfnamefont{M.}~\bibnamefont{Ave}} \bibnamefont{et~al.}
  (\bibinfo{collaboration}{Pierre Auger Collaboration}),
  \bibinfo{journal}{Nucl. Instrum. Meth.} \textbf{\bibinfo{volume}{A578}},
  \bibinfo{pages}{180} (\bibinfo{year}{2007}).

\bibitem[{\citenamefont{Abraham et~al.}(2010)}]{Abraham:2009pm}
\bibinfo{author}{\bibfnamefont{J.}~\bibnamefont{Abraham}} \bibnamefont{et~al.},
  \bibinfo{journal}{Nucl. Instrum. Meth.} \textbf{\bibinfo{volume}{A620}},
  \bibinfo{pages}{227} (\bibinfo{year}{2010}).

\bibitem[{\citenamefont{Tokuno et~al.}(2012)}]{Tokuno:2012mi}
\bibinfo{author}{\bibfnamefont{H.}~\bibnamefont{Tokuno}} \bibnamefont{et~al.},
  \bibinfo{journal}{Nucl. Instrum. Meth.} \textbf{\bibinfo{volume}{A676}},
  \bibinfo{pages}{54} (\bibinfo{year}{2012}).

\bibitem[{\citenamefont{Abreu et~al.}(2012{\natexlab{a}})}]{Abreu:2012pi}
\bibinfo{author}{\bibfnamefont{P.}~\bibnamefont{Abreu}} \bibnamefont{et~al.},
  \bibinfo{journal}{JINST} \textbf{\bibinfo{volume}{7}},
  \bibinfo{pages}{P10011} (\bibinfo{year}{2012}{\natexlab{a}}).

\bibitem[{\citenamefont{\ifmmode~\check{S}\else \v{S}\fi{}m\'{\i}da
  et~al.}(2014)}]{PhysRevLett.113.221101}
\bibinfo{author}{\bibfnamefont{R.}~\bibnamefont{\ifmmode~\check{S}\else
  \v{S}\fi{}m\'{\i}da}} \bibnamefont{et~al.}, \bibinfo{journal}{Phys. Rev.
  Lett.} \textbf{\bibinfo{volume}{113}}, \bibinfo{pages}{221101}
  (\bibinfo{year}{2014}).

\bibitem[{\citenamefont{{The Pierre Auger
  Collaboration}}(2015)}]{ThePierreAuger:2015rma}
\bibinfo{author}{\bibnamefont{{The Pierre Auger Collaboration}}},
  \bibinfo{journal}{Nucl. Instrum. Meth.} \textbf{\bibinfo{volume}{A798}},
  \bibinfo{pages}{172} (\bibinfo{year}{2015}).

\bibitem[{\citenamefont{Abu-Zayyad et~al.}(2012)}]{AbuZayyad201287}
\bibinfo{author}{\bibfnamefont{T.}~\bibnamefont{Abu-Zayyad}}
  \bibnamefont{et~al.}, \bibinfo{journal}{Nucl. Instrum. Meth.}
  \textbf{\bibinfo{volume}{A689}}, \bibinfo{pages}{87} (\bibinfo{year}{2012}).

\bibitem[{\citenamefont{Sokolsky}(2011)}]{Sokolsky201174}
\bibinfo{author}{\bibfnamefont{P.}~\bibnamefont{Sokolsky}},
  \bibinfo{journal}{Nuclear Physics B - Proceedings Supplements}
  \textbf{\bibinfo{volume}{212--213}}, \bibinfo{pages}{74}
  (\bibinfo{year}{2011}).

\bibitem[{\citenamefont{Chiba et~al.}(1992)}]{Chiba:1991nf}
\bibinfo{author}{\bibfnamefont{N.}~\bibnamefont{Chiba}} \bibnamefont{et~al.},
  \bibinfo{journal}{Nucl. Instrum. Meth.} \textbf{\bibinfo{volume}{A311}},
  \bibinfo{pages}{338} (\bibinfo{year}{1992}).

\bibitem[{Aab(2015)}]{Aab:2015bza}
\emph{\bibinfo{title}{{The Pierre Auger Observatory: Contributions to the 34th
  International Cosmic Ray Conference (ICRC 2015)}}} (\bibinfo{year}{2015}),
  \urlprefix\url{http://inspirehep.net/record/1393211/files/arXiv:1509.03732.pdf}.

\bibitem[{\citenamefont{Cazon et~al.}(2005)\citenamefont{Cazon, Vazquez, and
  Zas}}]{Cazon:2004zx}
\bibinfo{author}{\bibfnamefont{L.}~\bibnamefont{Cazon}},
  \bibinfo{author}{\bibfnamefont{R.}~\bibnamefont{Vazquez}}, \bibnamefont{and}
  \bibinfo{author}{\bibfnamefont{E.}~\bibnamefont{Zas}},
  \bibinfo{journal}{Astropart. Phys.} \textbf{\bibinfo{volume}{23}},
  \bibinfo{pages}{393} (\bibinfo{year}{2005}).

\bibitem[{\citenamefont{Cazon et~al.}(2004)\citenamefont{Cazon, Vazquez,
  Watson, and Zas}}]{Cazon:2003ar}
\bibinfo{author}{\bibfnamefont{L.}~\bibnamefont{Cazon}},
  \bibinfo{author}{\bibfnamefont{R.}~\bibnamefont{Vazquez}},
  \bibinfo{author}{\bibfnamefont{A.}~\bibnamefont{Watson}}, \bibnamefont{and}
  \bibinfo{author}{\bibfnamefont{E.}~\bibnamefont{Zas}},
  \bibinfo{journal}{Astropart. Phys.} \textbf{\bibinfo{volume}{21}},
  \bibinfo{pages}{71} (\bibinfo{year}{2004}).

\bibitem[{\citenamefont{Arsene et~al.}(2012)\citenamefont{Arsene, Rebel, and
  Sima}}]{Arsene}
\bibinfo{author}{\bibfnamefont{N.}~\bibnamefont{Arsene}},
  \bibinfo{author}{\bibfnamefont{H.}~\bibnamefont{Rebel}}, \bibnamefont{and}
  \bibinfo{author}{\bibfnamefont{O.}~\bibnamefont{Sima}}, \bibinfo{journal}{AIP
  Conf.Proc.} \textbf{\bibinfo{volume}{1498}}, \bibinfo{pages}{304}
  (\bibinfo{year}{2012}).

\bibitem[{\citenamefont{Arsene and Sima}(2015)}]{Arsene_Sima}
\bibinfo{author}{\bibfnamefont{N.}~\bibnamefont{Arsene}} \bibnamefont{and}
  \bibinfo{author}{\bibfnamefont{O.}~\bibnamefont{Sima}}, \bibinfo{journal}{AIP
  Conf.Proc.} \textbf{\bibinfo{volume}{1645}}, \bibinfo{pages}{286}
  (\bibinfo{year}{2015}).

\bibitem[{\citenamefont{Heck and Knapp}(1989)}]{corsika}
\bibinfo{author}{\bibfnamefont{D.}~\bibnamefont{Heck}} \bibnamefont{and}
  \bibinfo{author}{\bibfnamefont{J.}~\bibnamefont{Knapp}},
  \bibinfo{journal}{Report {\bf FZKA 6097} (1998), Forschungszentrum Karlsruhe;
  available from http://www-ik.fzk.de/\textasciitilde heck/publications/}
  (\bibinfo{year}{1989}).

\bibitem[{\citenamefont{Heck et~al.}(1998)\citenamefont{Heck, Knapp,
  Capdevielle, Schatz, and Thouw}}]{corsika1}
\bibinfo{author}{\bibfnamefont{D.}~\bibnamefont{Heck}},
  \bibinfo{author}{\bibfnamefont{J.}~\bibnamefont{Knapp}},
  \bibinfo{author}{\bibfnamefont{J.}~\bibnamefont{Capdevielle}},
  \bibinfo{author}{\bibfnamefont{G.}~\bibnamefont{Schatz}}, \bibnamefont{and}
  \bibinfo{author}{\bibfnamefont{T.}~\bibnamefont{Thouw}},
  \bibinfo{journal}{Report {\bf FZKA 6019} (1998), Forschungszentrum Karlsruhe;
  available from http://www-ik.fzk.de/corsika/physics$\_
  $description/corsika$\_ $phys.html \hspace{0.2em}}  (\bibinfo{year}{1998}).

\bibitem[{\citenamefont{Billoir}(2008)}]{Billoir:2008zz}
\bibinfo{author}{\bibfnamefont{P.}~\bibnamefont{Billoir}},
  \bibinfo{journal}{Astropart. Phys.} \textbf{\bibinfo{volume}{30}},
  \bibinfo{pages}{270} (\bibinfo{year}{2008}).

\bibitem[{\citenamefont{Wainberg et~al.}(2014)}]{Wainberg:2013koa}
\bibinfo{author}{\bibfnamefont{O.}~\bibnamefont{Wainberg}}
  \bibnamefont{et~al.}, \bibinfo{journal}{JINST} \textbf{\bibinfo{volume}{9}},
  \bibinfo{pages}{T04003} (\bibinfo{year}{2014}).

\bibitem[{\citenamefont{Videla et~al.}(2015)}]{Videla:2015xia}
\bibinfo{author}{\bibfnamefont{M.}~\bibnamefont{Videla}} \bibnamefont{et~al.},
  \bibinfo{journal}{Nucl. Instrum. Meth.} \textbf{\bibinfo{volume}{A791}},
  \bibinfo{pages}{6} (\bibinfo{year}{2015}).

\bibitem[{\citenamefont{Aab et~al.}(2016)}]{Aab:JINST2016}
\bibinfo{author}{\bibfnamefont{A.}~\bibnamefont{Aab}} \bibnamefont{et~al.}
  (\bibinfo{collaboration}{Pierre Auger Collaboration}),
  \bibinfo{journal}{{Prototype muon detectors for the AMIGA component of Pierre
  Auger Observatory, to appear in JINST}}  (\bibinfo{year}{2016}).

\bibitem[{\citenamefont{Aab et~al.}(2014{\natexlab{a}})}]{Aab:2014aea}
\bibinfo{author}{\bibfnamefont{A.}~\bibnamefont{Aab}} \bibnamefont{et~al.}
  (\bibinfo{collaboration}{Pierre Auger Collaboration}),
  \bibinfo{journal}{Phys. Rev.} \textbf{\bibinfo{volume}{D90}},
  \bibinfo{pages}{122006} (\bibinfo{year}{2014}{\natexlab{a}}).

\bibitem[{\citenamefont{Abreu
  et~al.}(2012{\natexlab{b}})}]{Collaboration:2012wt}
\bibinfo{author}{\bibfnamefont{P.}~\bibnamefont{Abreu}} \bibnamefont{et~al.}
  (\bibinfo{collaboration}{Pierre Auger Collaboration}),
  \bibinfo{journal}{Phys. Rev.Lett.} \textbf{\bibinfo{volume}{109}},
  \bibinfo{pages}{062002} (\bibinfo{year}{2012}{\natexlab{b}}).

\bibitem[{\citenamefont{Matthews}(2005)}]{Matthews:2005sd}
\bibinfo{author}{\bibfnamefont{J.}~\bibnamefont{Matthews}},
  \bibinfo{journal}{Astropart. Phys.} \textbf{\bibinfo{volume}{22}},
  \bibinfo{pages}{387} (\bibinfo{year}{2005}).

\bibitem[{\citenamefont{Pierog et~al.}(2013)\citenamefont{Pierog, Karpenko,
  Katzy, Yatsenko, and Werner}}]{Pierog:2013ria}
\bibinfo{author}{\bibfnamefont{T.}~\bibnamefont{Pierog}},
  \bibinfo{author}{\bibfnamefont{I.}~\bibnamefont{Karpenko}},
  \bibinfo{author}{\bibfnamefont{J.}~\bibnamefont{Katzy}},
  \bibinfo{author}{\bibfnamefont{E.}~\bibnamefont{Yatsenko}}, \bibnamefont{and}
  \bibinfo{author}{\bibfnamefont{K.}~\bibnamefont{Werner}}
  (\bibinfo{year}{2013}), \eprint{arXiv/1306.0121}.

\bibitem[{\citenamefont{Ferrari et~al.}(2005)\citenamefont{Ferrari, Sala,
  Fassò, and Ranft}}]{Ferrari:898301}
\bibinfo{author}{\bibfnamefont{A.}~\bibnamefont{Ferrari}},
  \bibinfo{author}{\bibfnamefont{P.~R.} \bibnamefont{Sala}},
  \bibinfo{author}{\bibfnamefont{A.}~\bibnamefont{Fassò}}, \bibnamefont{and}
  \bibinfo{author}{\bibfnamefont{J.}~\bibnamefont{Ranft}},
  \emph{\bibinfo{title}{{FLUKA: A multi-particle transport code (program
  version 2005)}}} (\bibinfo{publisher}{CERN}, \bibinfo{address}{Geneva},
  \bibinfo{year}{2005}), \urlprefix\url{https://cds.cern.ch/record/898301}.

\bibitem[{\citenamefont{Brancus et~al.}(2003)}]{Brancus}
\bibinfo{author}{\bibfnamefont{I.}~\bibnamefont{Brancus}} \bibnamefont{et~al.},
  \bibinfo{journal}{J.Phys.} \textbf{\bibinfo{volume}{G29}},
  \bibinfo{pages}{453} (\bibinfo{year}{2003}).

\bibitem[{\citenamefont{Haeusler et~al.}(2002)\citenamefont{Haeusler, Badea,
  Rebel, Brancus, and Oehlschlager}}]{Haeusler}
\bibinfo{author}{\bibfnamefont{R.}~\bibnamefont{Haeusler}},
  \bibinfo{author}{\bibfnamefont{A.}~\bibnamefont{Badea}},
  \bibinfo{author}{\bibfnamefont{H.}~\bibnamefont{Rebel}},
  \bibinfo{author}{\bibfnamefont{I.}~\bibnamefont{Brancus}}, \bibnamefont{and}
  \bibinfo{author}{\bibfnamefont{J.}~\bibnamefont{Oehlschlager}},
  \bibinfo{journal}{Astropart. Phys.} \textbf{\bibinfo{volume}{17}},
  \bibinfo{pages}{421} (\bibinfo{year}{2002}).

\bibitem[{\citenamefont{Rebel et~al.}(1995)\citenamefont{Rebel, Voelker,
  Foeller, and Chilingarian}}]{Rebel:1994ed}
\bibinfo{author}{\bibfnamefont{H.}~\bibnamefont{Rebel}},
  \bibinfo{author}{\bibfnamefont{G.}~\bibnamefont{Voelker}},
  \bibinfo{author}{\bibfnamefont{M.}~\bibnamefont{Foeller}}, \bibnamefont{and}
  \bibinfo{author}{\bibfnamefont{A.}~\bibnamefont{Chilingarian}},
  \bibinfo{journal}{J.Phys.} \textbf{\bibinfo{volume}{G21}},
  \bibinfo{pages}{451} (\bibinfo{year}{1995}).

\bibitem[{\citenamefont{Antoni et~al.}(2003)}]{Antoni:2003gd}
\bibinfo{author}{\bibfnamefont{T.}~\bibnamefont{Antoni}} \bibnamefont{et~al.},
  \bibinfo{journal}{Nucl. Instrum. Meth.} \textbf{\bibinfo{volume}{A513}},
  \bibinfo{pages}{490} (\bibinfo{year}{2003}).

\bibitem[{\citenamefont{Andringa et~al.}(2012)\citenamefont{Andringa, Cazon,
  Conceicao, and Pimenta}}]{Andringa:2011ik}
\bibinfo{author}{\bibfnamefont{S.}~\bibnamefont{Andringa}},
  \bibinfo{author}{\bibfnamefont{L.}~\bibnamefont{Cazon}},
  \bibinfo{author}{\bibfnamefont{R.}~\bibnamefont{Conceicao}},
  \bibnamefont{and} \bibinfo{author}{\bibfnamefont{M.}~\bibnamefont{Pimenta}},
  \bibinfo{journal}{Astropart. Phys.} \textbf{\bibinfo{volume}{35}},
  \bibinfo{pages}{821} (\bibinfo{year}{2012}).

\bibitem[{\citenamefont{Aab et~al.}(2014{\natexlab{b}})}]{Aab:2014dua}
\bibinfo{author}{\bibfnamefont{A.}~\bibnamefont{Aab}} \bibnamefont{et~al.}
  (\bibinfo{collaboration}{Pierre Auger Collaboration}),
  \bibinfo{journal}{Phys. Rev.} \textbf{\bibinfo{volume}{D90}},
  \bibinfo{pages}{012012} (\bibinfo{year}{2014}{\natexlab{b}}).

\bibitem[{\citenamefont{Sima et~al.}(2011)}]{Sima:2011zz}
\bibinfo{author}{\bibfnamefont{O.}~\bibnamefont{Sima}} \bibnamefont{et~al.},
  \bibinfo{journal}{Nucl. Instrum. Meth.} \textbf{\bibinfo{volume}{A638}},
  \bibinfo{pages}{147} (\bibinfo{year}{2011}).

\bibitem[{\citenamefont{Gaisser and Hillas}(1977)}]{GH}
\bibinfo{author}{\bibfnamefont{T.}~\bibnamefont{Gaisser}} \bibnamefont{and}
  \bibinfo{author}{\bibfnamefont{A.}~\bibnamefont{Hillas}},
  \bibinfo{journal}{Proc. of 15th ICRC 8 Plovdiv, Bulgaria}
  \textbf{\bibinfo{volume}{353}} (\bibinfo{year}{1977}).

\bibitem[{\citenamefont{Aab et~al.}(2015)}]{Aab:2014pza}
\bibinfo{author}{\bibfnamefont{A.}~\bibnamefont{Aab}} \bibnamefont{et~al.}
  (\bibinfo{collaboration}{Pierre Auger Collaboration}),
  \bibinfo{journal}{Phys. Rev.} \textbf{\bibinfo{volume}{D91}},
  \bibinfo{pages}{032003} (\bibinfo{year}{2015}).

\end{thebibliography}

\end{document}